# Intrinsic ferroelectric properties of strained tetragonal $PbZr_{0.2}Ti_{0.8}O_3$ obtained on layer-by-layer grown, defect-free single crystalline films**

By *Ionela Vrejoiu*, Gwenaël Le Rhun, Lucian Pintilie, Dietrich Hesse, Marin Alexe, and Ulrich Gösele*

[*] Dr. I. Vrejoiu, Dr. G. Le Rhun, Dr. L. Pintilie, Prof. D. Hesse, Dr. M. Alexe, Prof. U. Gösele

Max Planck Institute of Microstructure Physics, Weinberg 2, 06120 Halle, Germany

E-mail: vrejoiu@mpi-halle.de

[**] The work has been partly funded by Volkswagen Foundation, through the "Nanosized ferroelectric hybrids" project no I/80897. Sincere thanks are due to S. Reyntjens and E. Sourty, both from FEI Application Laboratory (Eindhoven/NL), for the FIB preparation and STEM investigation, respectively, to N. D. Zakharov for HR-TEM investigations, to S. Swatek for the TEM sample preparation, and N. Schammelt for assistance in PLD-system maintenance.

In many materials of technological interest, especially in semiconductors, structural defects play detrimental roles.[1] In case of silicon (and other semiconductors) defects have to be thoroughly controlled. The growth of defect-free single crystals was a prerequisite for the accurate assessment of intrinsic physical properties of silicon, such as electronic band structure, optical bandgap, carrier densities, etc. Defects are not always associated with negative changes of properties. Some defects are desirable, such as the doping of a semiconductor with foreign atoms, which can alter the electrical conductivity by many orders of magnitude in a controlled manner.[2] In the case of silicon homoepitaxy, the growth of silicon on silicon surfaces is of interest both from a technological and a scientific point of view.

Here we report on the intrinsic physical properties of strained, defect-free, single crystalline $PbZr_{0.2}Ti_{0.8}O_3$ films synthesized by pulsed-laser deposition (PLD). $PbZr_xTi_{1-x}O_3$ (PZT) ferroelectrics are considered for most applications involving ferroelectric oxide materials, from piezoelectric transducers to dynamic nonvolatile random access memories. The incongruent melting of $PbZrO_3$ and $PbTiO_3$ as well as the high volatility of PbO at growth temperatures pose a challenge in establishing the equilibrium of bulk crystals of PZT.[3] Therefore, single-domain single crystals of PZT have never been synthesized for a significant compositional range across the solid-solution phase. This leaves the intrinsic properties of single domain PZT crystals under debate.[4] Hence, the growth of single crystalline and defect-free PZT thin films may enable the study of fundamental physical properties. For instance, the spontaneous polarization $P_S$, which is a fundamental parameter that defines the performance of a ferroelectric material, is rarely reported due to the lack of single crystals[4] and perfectly polar-axis-oriented films.[5,6] Although the macroscopic polarization is the basic quantity used to describe ferroelectrics, it has long escaped precise microscopic definition. It has been pointed out that the previously calculated values[4] of the polarization may thus be incorrect, due to invalid definitions and models employed for their computation.[7] Novel concepts such as polarization as a Berry phase of the electronic Bloch wavefunctions have been recently introduced.[8,9]

Additionally, accurate investigations of the highly debated issue of size-effects in nanostructured ferroelectrics demand the growth of defect-free materials, to rule out the extrinsic contribution of defects to the disappearance/suppression of ferroelectricity below a certain critical size.[10,11]

The fabrication of defect-free single crystalline ferroelectric thin films requires the careful choice of a single crystalline substrate and a growth technique. It is necessary to employ a single crystalline substrate with closely matched lattice constants. In mismatched epitaxial films which grow in a two-dimensional mode (i.e., step flow or layer-by-layer growth) biaxial stress is generated. Usually the stress relaxation in mismatched epitaxial films leads to the formation of misfit dislocations (MDs) at the film/substrate interface[12-14], which are commonly accompanied by threading dislocations (TDs) that extend across the film.[12] It has been demonstrated that MDs have an important impact on the ferroelectric properties of nanostructured ferroelectric perovskites.[15,16]

Pulsed-laser deposition (PLD) is suitable for synthesis of high quality homo- and heteroepitaxial films. This is due to the highly supersaturated ablation plasma plume, its pulsed nature and the adjustable deposition rate attainable by changing the laser energy density, laser repetition rate, background reactive gas pressure, and target-to-substrate distance.[17-19]

Bearing in mind the aforementioned reasoning, we chose to grow epitaxial PZT films onto vicinal $SrTiO_3$ (001) (STO) substrates by PLD. First $SrRuO_3$ (SRO) was deposited on STO as bottom electrode, also by PLD. SRO is an excellent template for the heteroepitaxial growth of high quality ferroelectric perovskites, as it grows in a single crystalline manner and with an atomically flat surface on vicinal STO (001) substrates.[20] Concerning PZT, we focused on the $PbZr_{0.2}Ti_{0.8}O_3$ composition, since it has an in-plane lattice parameter of 3.935 Å, which is closely matched to that of SRO (3.928 Å). Microstructural investigations, macroscopic measurements of ferroelectric polarization and dielectric constant, and local measurements of the piezoelectric coefficient were performed to characterize the single crystalline PZT films.

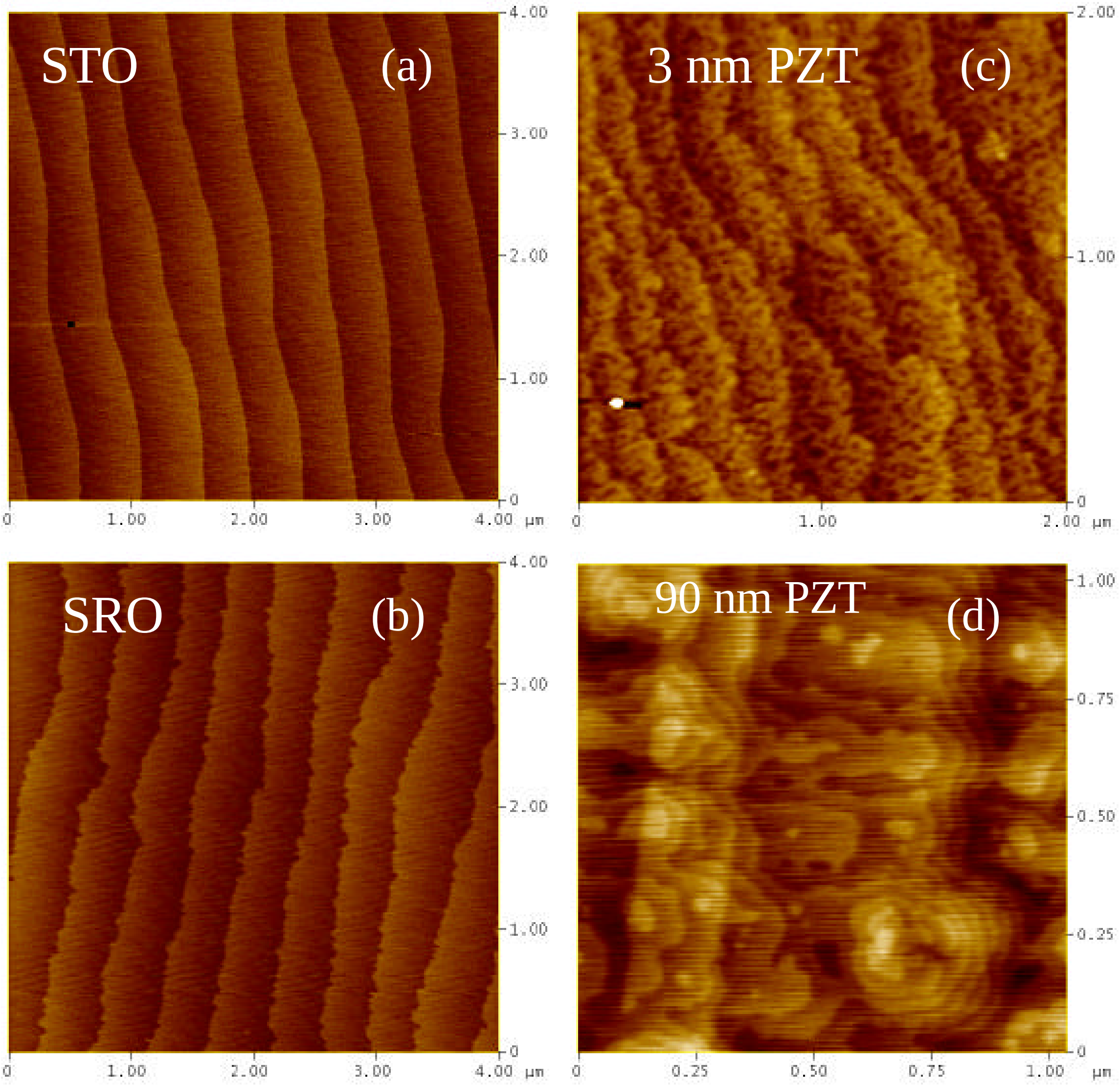

**Figure 1**. AFM images of the vicinal $SrTiO_3$ substrate and the eptaxially PLD-grown $SrRuO_3$ and $PbZr_{0.2}Ti_{0.8}O_3$ layers. The one-unit-cell stepped terraces (400 nm width) of a $SrTiO_3$ substrate are visible in (a). The morphology of the $SrRuO_3$ step-flow grown (t= 20 nm) on this substrate is shown in (b). In (c) and (d) the morphology of layer-by-layer grown ultrathin (t≈ 3 nm) and thick (t= 90 nm) $PbZr_{0.2}Ti_{0.8}O_3$ films deposited on top of $SrRuO_3/SrTiO_3$ heterostructures are displayed.

The surface morphology of the layers was investigated by atomic force microscopy (AFM). The SRO layer grew atomically flat, with one unit-cell stepped terraces (step-flow growth regime) (Fig. 1(b)), following the terraces of the vicinal STO substrate (Fig. 1(a)).[20] It was previously pointed out that the flatness of STO and SRO surfaces plays an important role in the growth of ultrathin PZT films.[21] AFM images taken on the surface of an ultrathin PZT film (film thickness t≈ 3 nm) reveal that, at the initial stage of the growth, the PZT layer grows by filling the terraces of the template SRO layer (Fig. 1(c)), and layer-by-layer growth[22] prevails also for much thicker PZT layers (t≈ 90 nm), as shown in Fig.1(d). In the layer-by-layer growth regime, the indications are that epitaxy results in almost all circumstances provided that the substrate surface is clean enough and well prepared. In this regime the substrate has a very strong influence on the form of the film produced, and the growing film has little option but to choose the best (i.e., necessarily epitaxial) orientation in which to grow.[22]

Cross-sectional transmission electron microscopy (TEM), atomic number (Z)-contrast scanning transmission electron microscopy (STEM), and electron diffraction investigations revealed the entire heterostructure to be epitaxial, and the interfaces between the layers to be plane and atomically sharp (Fig. 2(a), (c), (d)). TEM allowed to measure the layer thicknesses, i.e. 20 nm for the SRO layer and 90 –100 nm for the PZT layer (Fig. 2 (a)). 90° domains, which are typical for tetragonal perovskite ferroelectric films, were not formed in the PZT layer. This indicates that the PZT layer is strained and the strain was not relieved by the formation of twins.[23,24]

Polydomain formation in epitaxial films undergoing a phase transformation is a mechanism that relaxes the total strain energy, which is the result of lattice misfit, the difference in the thermal expansion coefficients of the film and the substrate, and misfit dislocation formation.

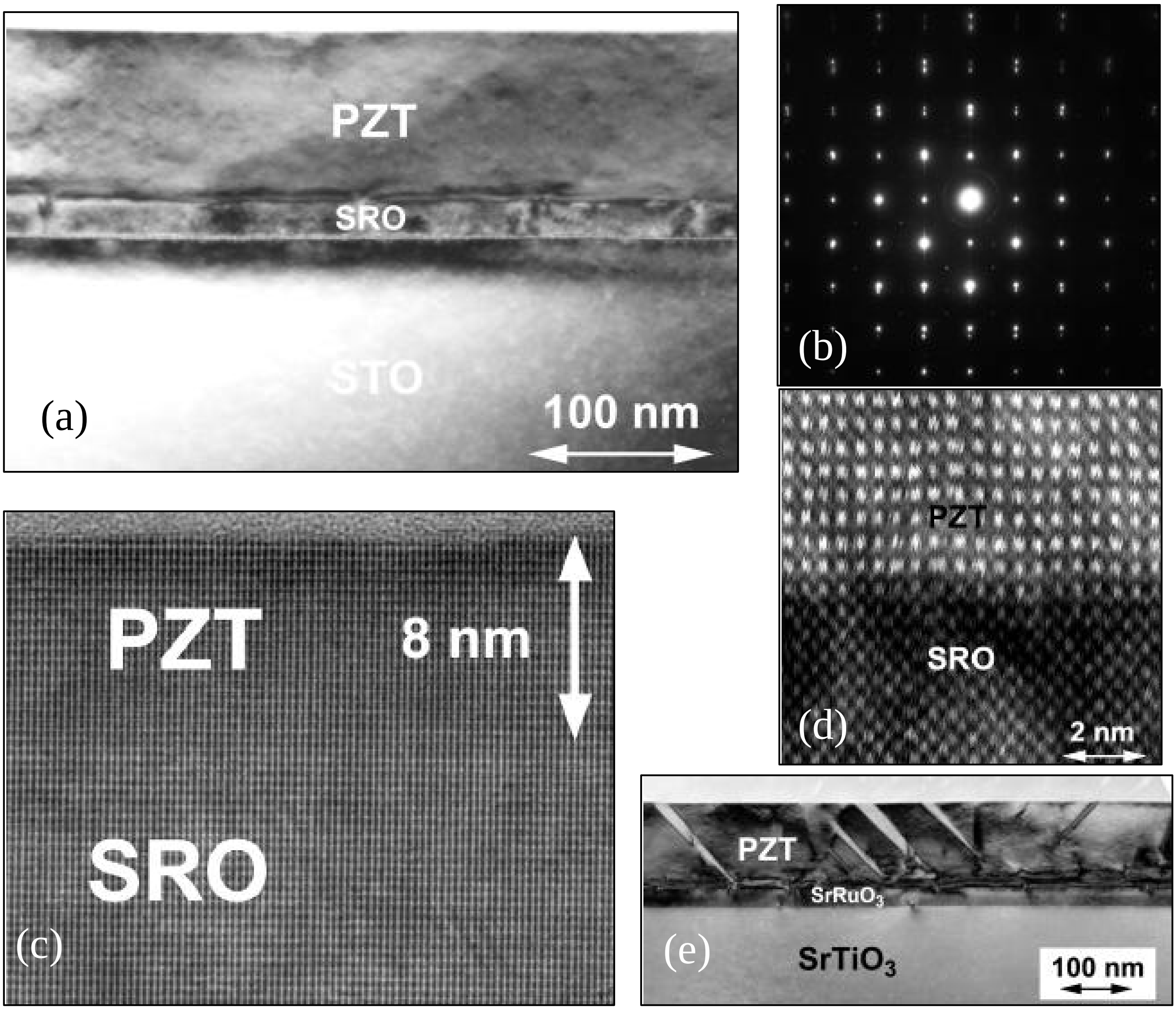

**Figure 2**. Transmission electron microscopy investigations of the $PbZr_{0.2}Ti_{0.8}O_3$/ $SrRuO_3$/$SrTiO_3$ heterostructures. The images are: (a) cross-section TEM micrograph of a defect-free PZT film (t= 90 nm) on top of a SRO film (t= 20 nm) grown on vicinal STO (001) and (b) electron-diffraction pattern of this heterostructure; (c) HRTEM micrograph of an ultrathin PZT film (8nm); (d) Z-contrast STEM

micrograph showing the sharp and dislocation-free interface between the PZT and SRO layers of such a heterostructure; (e) cross-section TEM micrograph of a defective PZT film (t=120 nm) on top of a SRO film (t= 20 nm) grown on vicinal STO (001).

High resolution TEM (HRTEM) of an ultrathin PZT film (8-9 nm) grown also in the layer-by-layer mode reveals that the PZT layer is perfectly strained to match the SRO lattice parameter (Fig 2(c)). No misfit dislocations were formed in the PZT layer. High resolution STEM Z-contrast micrographs of a 30 nm thick PZT film confirmed this very good matching and the absence of misfit dilocations, as shown in Fig 2(d). Hence, we were able to grow defect-free films from a few to several hundreds of nanometers thickness.

For comparison, we show also the TEM cross-section micrograph of an epitaxial PZT film of about the same thickness that has structural defects. This layer was synthesized with the same PLD parameters using a $Pb_{1.1}(Zr_{0.2}Ti_{0.8})O_3$ target that we suspect to be slightly oxygen-deficient (Fig 2(e)). This PZT layer has a considerably more defective microstructure and exhibits 90° domains. At the SRO/PZT interface, a defective thin layer was formed, that was evidenced by HRTEM as well (not shown here).

From the electron diffraction pattern shown in Fig 2(b), the tetragonality ratio of the defect-free PZT layer could be calculated, $c/a \approx 1.06$. This is considerably higher than that of bulk PZT with this composition, whose room temperature value is $c/a = 1.051$.[25] Recently Morioka *et al.*[5] reported on the relationship between the spontaneous polarization $P_S$ and $c/a$-1, for tetragonal PZT. From their data, a value of $P_S \approx 110\ \mu C/cm^2$ would result for $c/a \approx 1.058$.

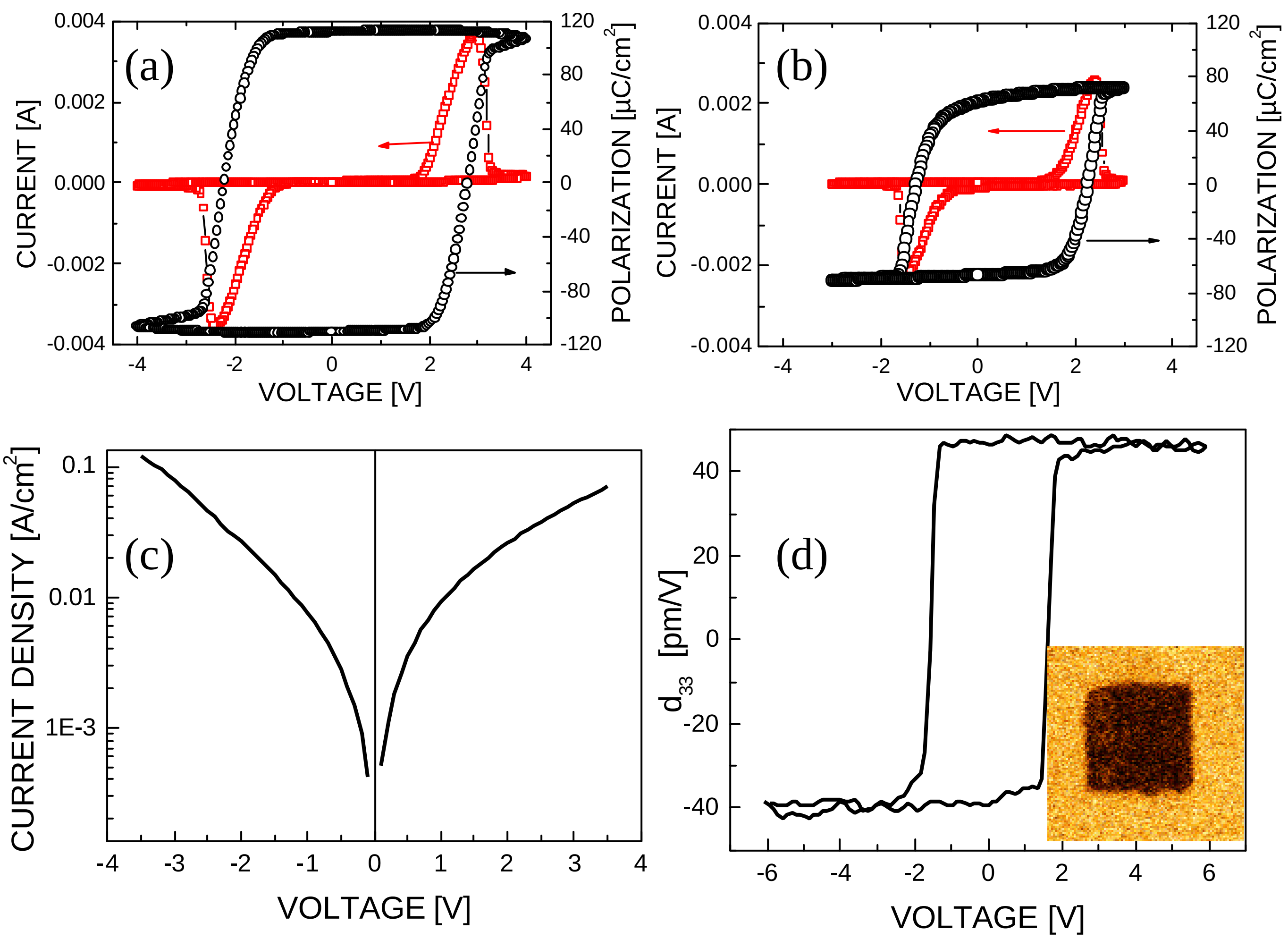

**Figure 3**. Ferroelectric and piezoelectric investigations of the $PbZr_{0.2}Ti_{0.8}O_3$/ $SrRuO_3$/$SrTiO_3$ heterostructures**.** Ferroelectrc hysteresis curves measured at 1 kHz are shown for (a) a defect-free PZT film ($P_r$= 105 ± 5 μC/cm$^2$) and (b) a defective film ($P_r$= 65 μC/cm$^2$). Leakage current density versus voltage measured on the defect-free PZT layer is shown in (c) and local piezo-coefficient hysteresis of the same layer is displayed in (d). The inset in (d) is a 1 μm x 1μm piezo-response image, where the center area (0.5 μm x 0.5μm) has switched polarization by being scanned with the tip under d.c. bias. No backswitching was observed after a time as long as 71 hours.

The ferroelectric and piezoelectric properties of the PZT layers were investigated in detail. Ferroelectric polarization and switching current hysteresis curves are shown in Figure 3(a). The polarization hysteresis curve has a square shape, remnant polarization being as high as $P_r \approx 105 \pm 5$ µC/cm$^2$, and basically equal to the spontaneous polarization, $P_S$. The polarization showed only a weak frequency dependence (within 10 %). This value is, to the best of our knowledge, the highest value reported for PZT epitaxial films with the considered composition.[5,11,26-28] It exceeds also the value of the remnant polarization recently reported for single *c*-domain heteroepitaxial $PbTiO_3$ films.[29] It is also higher than the theoretical value, $P_r = 70$ µC/cm$^2$, calculated for bulk single crystals.[4] It is, nevertheless, in very good agreement with the value of the tetragonality ratio mentioned above. We mention here that recent calculations[30] for $PbTiO_3$ that undergoes anomalous enhancement of tetragonality predict values of the spontaneous polarization as high as $P_S \approx 140$ µC/cm$^2$.

The hysteresis loop measured under the same conditions on the PZT layer with structural defects shown in Fig. 2(e) is displayed in Fig. 3(b) for comparison. The shape is less square and the remnant polarization much lower, $P_r \approx 65$ µC/cm$^2$. This should be a consequence of the 90° domains and the more defective microstructure of this epitaxial PZT layer.

Figure 3(c) displays the leakage current density for the defect-free layer. The slight asymmetry of the two branches of the curve indicates that interface-controlled injection of carriers occurs. Even though both the bottom and the top electrodes consisted of SRO, they were deposited under different conditions, and therefore the interface may have different properties, namely different interface state densities and potential barrier heights. The leakage current density has a rather large value, especially when compared with the leakage currents measured in polycrystalline PZT films. We suppose that in defect-free single-crystalline films the carrier mobility is much higher than in a defective film. Polycrystalline films exhibit extended defects (i.e., grain boundaries, domains) that may scatter the injected carriers, thus reducing their mobility, and may trap the carriers decreasing thus their density.[31] Nevertheless, we should mention that the relatively high leakage current density of our defect-free single crystalline PZT films does not hinder in any way the ferroelectric switching.

The dielectric constant was estimated from capacitance measurements. The obtained value was around $\varepsilon_{33}= 90 \pm 5$, which is significantly lower than the values reported by other groups for *c*-axis epitaxial PZT films.[6,28,32] However, it is in good agreement with a theoretical value that Huan *et al*.[4] estimated for bulk single crystals of PZT with this composition. For the defective layer that has 90° domains (Fig. 2(e)), based on the capacitance measurements, we estimated the effective dielectric permittivity to be about 168. This is an indication that the value we measured for the defect-free PZT layer, $\varepsilon_{33}= 90 \pm 5$, is the intrinsic value of dielectric permittivity, as there are no extended defects, such as 90° domains and grain boundaries, that may extrinsically enhance the measured dielectric permittivity.[33]

Local piezoresponse behavior was investigated by a scanning probe microscope. The investigations revealed that the PZT layer has an out-of-plane polarization that is preferentially oriented in the as-grown state[34,35] and confirmed the absence of 90° domains, as it can be seen in the inset of Fig 3(d). The piezoresponse hysteresis curve plotted in Fig. 3(d) has a square shape and the value of the $d_{33}$ piezoelectric coefficient was estimated to be around $d_{33}= 45 \pm 5$ pm/V[4,35], that is also a theoretical value for $PbZr_{0.2}Ti_{0.8}O_3$ single-domain strained films on $SrTiO_3$.[27] The inset in Fig. 3(d) shows the piezoresponse measured 71 hours after poling the center of the scanned region (1 x 1 μm$^2$) with negative bias (-5 V), to switch the polarization direction. No backswitching of the reversed polarization to the as-grown polarization state occurred.

In conclusion, we have synthesized and studied defect-free single crystalline $PbZr_{0.2}Ti_{0.8}O_3$ thin films on SRO-coated STO (001) vicinal single crystals. The PZT films have square shape polarization and piezoelectric hysteresis loops, remnant polarization values of up to $P_r$= 105 ± 5 μC/cm$^2$, a dielectric constant $\varepsilon_{33}$= 90 ± 5, and a piezoelectric coefficient of up to $d_{33}$= 45 ± 5 pm/V. Both macroscopic ferroelectric measurements and local piezoresponse investigations point out a remarkable ease of polarization switching, that is likely to be the consequence of the lack of defects in the PZT layer. Synthesis of defect-free single crystalline films enabled us to unambiguously determine fundamental intrinsic quantities that describe ferroelectric materials, such as spontaneous polarization and dielectric constant, and will allow a more reliable study of size effects in defect-free nanostructured ferroelectrics.

EXPERIMENTAL

Vicinal single crystalline STO (001) substrates with a miscut angle of 0.05°-0.2° (CrysTec, Berlin) were used to epitaxially grow the heterostructure. The STO substrates were etched in a buffered HF solution and annealed in air at temperatures between 950°C and 1100°C, depending on the miscut angle.[19,20] Thus one unit cell-stepped terraces with straight ledges formed. The layers were fabricated by PLD, employing a KrF excimer laser (λ= 248 nm). Ceramic $SrRuO_3$ and $Pb_{1.1}(Zr_{0.2}Ti_{0.8})O_3$ targets (PRAXAIR) were used for PLD. The SRO layer (t= 20-100 nm) was deposited at a substrate temperature T= 700°C in a background atmosphere of 100 mTorr oxygen, with a laser fluence $\Phi_L$= 1.5 - 2 J/cm$^2$, at a laser repetition rate $\nu_L$= 5 Hz. The subsequent PZT layer (t= 2 - 275 nm) was grown at T= 575°C in 200 mTorr oxygen, $\Phi_L$= 2 - 3 J/cm$^2$, and at $\nu_L$ = 3 or 5 Hz. Circular SRO top-electrodes (pads of 340 µm diameter) were deposited by PLD at room temperature through a shadow-mask and then platinum was deposited by sputtering on top of the SRO, in order to ease the contacting of the capacitor pads.

TEM samples were prepared by standard mechanical and ion-beam thinning procedures.[36] The cross-section sample used for Z-contrast STEM investigations of the interfaces was prepared by focused ion beam technique. TEM was performed in a Philips CM20T microscope, and STEM in a FEI Tecnai G$^2$F20 Xtwin microscope, both operated at 200 kV.

Macroscopic ferroelectric hysteresis, fatigue and leakage current curves were acquired[37] using the TF2000 Analyzer (AixaCCT). The capacitance measurements were performed at zero dc voltage and with an amplitude of 50 mV for the 1 kHz ac probe signal.

Local piezoresponse behavior was investigated by a scanning probe microscope (ThermoMicroscope). PtIr coated tips (Nanosensors, ATEC-EFM) with an elastic constant of about 2.5 N/m were employed. Local piezoelectric hystersis curve was acquired by superimposing a d.c. bias voltage to the a.c. probing voltage (1 V, 22.3 kHz). For quantitative measurements the piezoresponse signal was previously calibrated using an x-cut quartz ($d_{33}$= 2.17 pm/V).

**References**

[1] S. K. Estreicher, *Mater. Today* **2003**, *6*, 26.

[2] H. J. Queisser, E. E. Haller, *Science* **1998**, *281*, 945.

[3] S. Aggarwal, R. Ramesh, *Annu. Rev. Mater. Sci.* **1998**, *28*,463.

[4] M. J. Haun, E. Furman, S. J. Jang, L. E. Cross *Ferroelectrics* **1989**, *99*, 63.

[5] H. S. Morioka, S. Yokoyama, T. Oikawa, H. Funakubo, K. Saito, *Appl. Phys. Lett.* **2004**, *85*, 3516.

[6] Y. K. Kim, H. Morioka, R. Ueno, S. Yokoyama, H. Funakubo, *Appl. Phys. Lett.* **2005**, *86*,212905.

[7] D. Vanderbilt, R. D. King-Smith ,*Phys. Rev. B* **1993**, *48*,4442.

[8] R. Resta, *Europhys. News* **1997**, *28*,18.

[9] N. Sai, K. M. Rabe, D. Vanderbilt, *Phys. Rev. B* **2002**, *66*,104108.

[10] J. Junquera, P. Ghosez,. *Nature* **2003**, *422*, 506.

[11] V. Nagarajan, S. Prasertchoung, T. Zhao, H. Zheng, J. Ouyang, R. Ramesh, W. Tian, X. Q. Pan, D. M. Kim, C. B. Eom, H. Kohlstedt, R. Waser, *Appl. Phys. Lett.* **2004**, *84*, 5225.

[12] E. A. Fitzgerald, *Mater. Sci. Rep.* **1991**, *7*, 87.

[13] A. Y. Emelyanov, N. A. Pertsev, *Phys. Rev. B* **2003**, *68*,214103.

[14] S. Y. Hu, Y. L. Li, L. Q. Chen, *J. Appl. Phys.* **2003**, *94*, 2542.

[15] M.-W. Chu, I. Szafraniak, R. Scholz, C. Harnagea, D. Hesse, M. Alexe, U. Gösele,. *Nature Mater.* **2004**, *3*, 87.

[16] V. Nagarajan, C. J. Lia, H. Kohlstedt, R. Waser, I. B. Misirlioglu, S. P. Alpay, and R. Ramesh, Appl. Phys. Lett. **2005**, *86*,192910.

[17] D. Bäuerle, *Laser Processing and Chemistry*, 3rd ed., **2000**, Springer, New York

[18] P. R. Willmott, J. R. Huber, *Rev. Mod. Phys.* **2000**, *72*, 315.

[19] G Koster, *Artificially Layered Oxides by Pulsed Laser Deposition, Ph. D. Thesis*, **1999**, University of Twente.

[20] W. Hong, H. N. Lee, M. Yoon, H. M. Christen, D. H. Lowndes, Z. Suo, Z. Zhang, *Phys. Rev. Lett.* **2005**, *95*,095501.

[21] H. Nonomura, H. Fujisawa, M. Shimizu, H. Niu, *Jpn. J. Appl. Phys*. **2002**, *41*,6682.

[22] J. W. Matthews, *Epitaxial Growth Part B.* **1975**, Academic Press, New York

[23] A. L. Roitburd, *Phys. Stat. Sol. (a)* **1976**, *37,* 329.

[24] V. Nagarajan, I. G. Jenkins, S. P. Alpay, H. *Li, S. Aggarwal, L. Salmanca-Riba, A. L. Roytburd, R. Ramesh, J. Appl. Phys. **1999**, 86*, 595.

[25] Y. Li, V. Nagarajan, S. Aggarwal, R. Ramesh, L. G. Salamanca-Riba, and L. J. Martinez-Miranda, *J. Appl. Phys*. **2002**, *92*, 6762.

[26] V. Nagarajan, A. Roytburd, A. Stanishevsky, S. Prasertchoung, T. Zhao, L. Chen, J. Melngailis, O. Auciello, and R. Ramesh, *Nat. Mater*. **2002***, 2*, 43.

[27] N. A. Pertsev, V G. Kukhar, H. Kohlstedt, R. Waser, *Phys. Rev. B* **2003**, *67*, 054107.

[28] C. M. Foster, G. –R. Bai, R. Csencsits, J. Vetrone, R. Jammy, L. A. Willis, E. Carr, J. Amano, *J. Appl. Phys*. **1997**, *81*, 2349.

[29] W.W. Jung, H. C. Lee, W. S. Ahn, S. H. Ahn, S. K. Choi, *Appl. Phys. Lett*. **2005**, *86*, 252901.

[30] S. Tinte, K. M. Rabe, D. Vanderbilt, *Phys. Rev. B* **2003**, *68*, 144105.

[31] L. Pintilie, M. Alexe, *J. Appl. Phys*., in press

[32] S. Yokoyama, Y. Honda, H. Morioka, S. Okamoto, H. Funakubo, T. Iijima, H. Matsuda, K. Saito, T. Yamamoto, H. Okino, O. Sakata, S. Kimura, *J. Appl. Phys*. **2005**, *98*, 094106.

[33] F. Xu, S. Trolier-McKinstry, W. Ren, B. Xu, X. –L. Xie, J. Hemker, *J. Appl. Phys*. **2001**, *89*,1336.

[34] C. S. Ganpule, V. Nagarajan, H. Li, A. S. Ogale, D. E. Steinhauer, S. Aggarwal, E. Williams, R. Ramesh, P. De Wolf, *Appl. Phys. Lett*. **2000**, *77*, 292.

[35] G. Le Rhun, I. Vrejoiu, L. Pintilie, D. Hesse, M. Alexe, U. Gösele, submitted to *Nat. Mater*.

[36] D. B. Williams, C. B. Carter, *Transmission Electron Microscopy*, **1996**,Plenum Press, New York

[37] L. Pintilie, I. Vrejoiu, D. Hesse, M. Alexe, accepted to *Appl. Phys. Lett.*